\documentclass[aps, 10pt, prb, showkeys, twocolumn, superscriptaddress]{revtex4-2}
\usepackage{graphicx, xcolor, amssymb, amsmath, amsfonts, enumitem}
\usepackage[utf8]{inputenc}
\usepackage[normalem]{ulem}
\usepackage[colorlinks=true, linkcolor=blue, citecolor=blue, urlcolor=blue]{hyperref}
\usepackage{url, soul, ulem, textgreek, bm, comment}
\usepackage[dvipsnames]{xcolor}
\usepackage{appendix}
\usepackage{kotex}
\usepackage{float}
\usepackage{orcidlink}
\usepackage{booktabs}

\newcommand{\apctpadd}{Asia Pacific Center for Theoretical Physics, Pohang, Gyeongbuk, 37673, Republic of Korea}
\newcommand{\postechadd}{Department of Physics, Pohang University of Science and Technology, Pohang, Gyeongbuk 37673, Republic of Korea}

\begin{document}

\title{Giant orbital Hall effect from cubic Dresselhaus orbital coupling}

\author{Gwen Sevilen\orcidlink{0009-0008-6771-9040}}
\affiliation{\mbox{Department of Physics, Institut Teknologi Bandung, Bandung, West Java 40132, Republic of Indonesia}}

\author{Kyoung-Min Kim\orcidlink{0000-0003-2468-3152}}
\email{kyoungmin.kim@apctp.org}
\affiliation{\apctpadd}
\affiliation{\postechadd}

\date{\today}

\begin{abstract}    
    The orbital Berry curvature (OBC) governs the intrinsic orbital Hall conductivity (OHC), a central quantity in orbitronics.
    Previous approaches for enhancing the OHC have primarily relied on a linear-in-momentum, Rashba-type orbital coupling.
    Here, we show that cubic Dresselhaus orbital coupling offers a new route to enhancing the OHC.
    Using an effective two-orbital band model, we show that the cubic coupling generates momentum-space hot spots, absent in the purely linear case, at which the OBC is strongly enhanced.
    This local enhancement, together with the multiplicity of the hot spots, boosts the OHC by more than an order of magnitude relative to the linear-coupling value.
    We further find that the OHC diverges inversely with the level splitting in the small-splitting limit, with a divergence coefficient universally seventeen times larger than that of the linear case.
    These results establish cubic Dresselhaus coupling as a route to giant orbital Hall responses, opening new avenues for orbitronic device applications.
\end{abstract}


\maketitle

\textit{Introduction.}---The field of orbitronics aims to harness the orbital angular momentum of Bloch electrons as an active degree of freedom for charge-based information technology~\cite{Go_2021,BurgosAtencia2024}.
The recognition that orbital angular momentum can be finite even in centrosymmetric, weakly spin--orbit-coupled solids~\cite{PhysRevLett.107.156803,PhysRevLett.128.176601} has spurred interest in orbital analogs of spintronic phenomena, most notably the orbital Hall effect (OHE)~\cite{PhysRevLett.95.066601,PhysRevLett.102.016601,PhysRevLett.100.096601}.
Since the OHE requires no spin-orbit coupling, unlike its spin counterpart, it survives in light metals~\cite{Go2018,Seifert2023,Go2023}, widening the material platform well beyond the heavy elements traditionally required.
This has motivated a growing body of experimental reports of sizable OHE responses in light and $3d$/$4d$ transition metals~\cite{Choi2023,Lyalin2023,PhysRevResearch.5.023054,10.1063/5.0293680}, together with demonstrations that the resulting orbital currents can drive current-induced torques and even magnetization switching~\cite{Ding2020,PhysRevResearch.2.013177,PhysRevResearch.2.013127,Moriya2024,Yang2024,vijayan2025}.

On the theoretical side, a series of works has established the microscopic foundations of orbital transport.
The intrinsic OHE was first predicted in $p$-doped silicon~\cite{PhysRevLett.95.066601} and in $4d$/$5d$ transition metals~\cite{PhysRevLett.100.096601,PhysRevLett.102.016601}, with orbital texture subsequently identified as its microscopic origin~\cite{Go2018}.
This picture was later extended to surface and interfacial settings~\cite{Go2017}, culminating in a recent universal formalism showing that, in an effective two-band model, the orbital Berry curvature (OBC) is fixed entirely by the ordinary Berry curvature and the band splitting~\cite{Lee2025}.
Notably, all of these studies build on the same underlying ingredient---a linear-in-momentum, Rashba-type orbital coupling.
One conceivable extension is to consider higher-order couplings, which have long been recognized as an important refinement in the analogous spin context~\cite{PhysRevLett.98.226802}.
This possibility has remained unexplored, however, and a concrete, Hamiltonian-level design principle for boosting the orbital Hall conductivity (OHC) beyond this linear paradigm has so far been lacking.

\begin{figure}[t!]
    \centering
    \includegraphics[width=1.0 \linewidth]{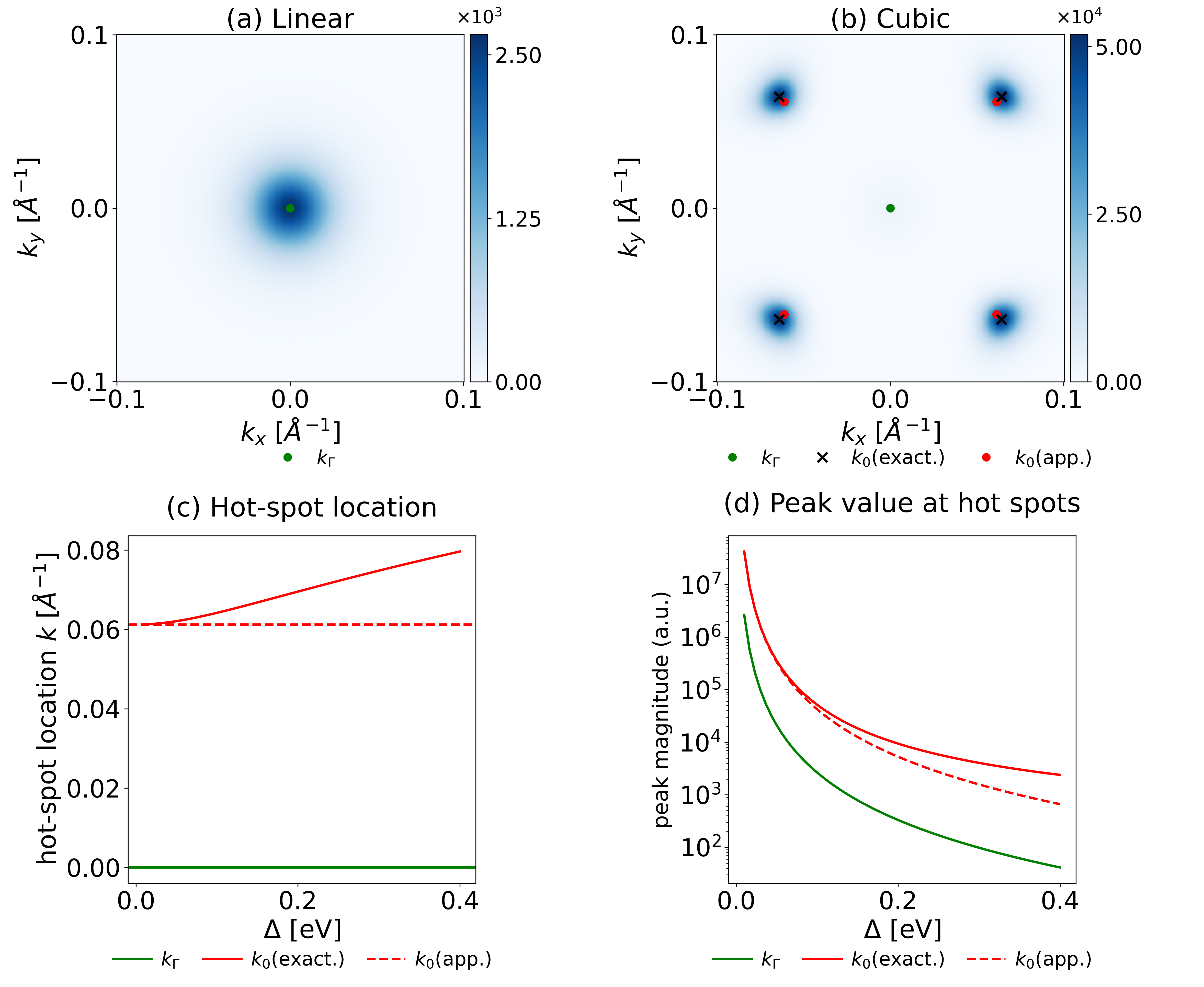}
    \caption{\textbf{Orbital Berry curvature (OBC).}
    (a)--(b) Momentum-space distribution of the OBC for (a) the linear case ($\beta=3~\mathrm{eV\,\AA}$, $\alpha=0$, and $\Delta=0.1~\mathrm{eV}$) and (b) the cubic case ($\beta=3~\mathrm{eV\,\AA}$, $\alpha=800~\mathrm{eV\,\AA^3}$, and $\Delta=0.1~\mathrm{eV}$).
    Green dots indicate the $\Gamma$ point, while red dots indicate the cubic Dresselhaus hot spots $\bm{k}_0$.
    (c) Hot-spot location as a function of the energy splitting $\Delta$ for $\beta=3~\mathrm{eV\,\AA}$ and $\alpha=800~\mathrm{eV\,\AA^3}$. The red solid line shows the numerically obtained values, while the red dashed line shows the approximate values $k_0=\sqrt{\beta/\alpha}$; the green line indicates the $\Gamma$ point.
    (d) Maximum value of the OBC at the hot spots for the same parameters as in (c). The solid line shows the numerically obtained values, while the dashed line shows the approximate value [see Eq.~\eqref{eq:orbital_berry_curvature_k0}]; the green line indicates the peak value at the $\Gamma$ point.
    }
    \label{fig1}
\end{figure}

In this work, we introduce ``cubic Dresselhaus orbital coupling'' as such a Hamiltonian-level lever for controlling the OHC.
Using an effective two-band orbital model, we show that the cubic term generates momentum-space hot spots, absent in the purely linear case, at which the OBC is locally maximized.
This local enhancement propagates into an order-of-magnitude enhancement of the OHC.
We further show that, in the small-splitting regime, the OHC diverges as $1/\Delta$, with a universal enhancement factor seventeen times stronger than in the linear case.
We finally demonstrate that the OHC exhibits characteristic temperature and thickness dependences, providing a decisive experimental signature of the cubic enhancement.

\textit{Model.}---We consider a two-dimensional system of two spinless orbital states described by the Hamiltonian
\begin{equation} \label{eq:hamiltonian}
\hat{H} = \sum_{\bm{k}} \begin{pmatrix} c_{\bm{k}a}^\dagger & c_{\bm{k}b}^\dagger \end{pmatrix} \begin{pmatrix} \epsilon_{\bm{k}}^a & \gamma_{\bm{k}}\\ \gamma_{\bm{k}}^* & \epsilon_{\bm{k}}^b \end{pmatrix} \begin{pmatrix} c_{\bm{k}a} \\ c_{\bm{k}b} \end{pmatrix},
\end{equation}
where $c_{\bm{k}o}^{\dagger}$ ($c_{\bm{k}o}$) creates (annihilates) an electron with momentum $\bm{k}$ in orbital state $o=a,b$. The energy dispersions for the two orbital states are given by
\begin{equation} \label{eq:energy_dispersion}
\epsilon_{\bm{k}}^a = \frac{\hbar^2k^2}{2m^*} + \Delta, \qquad \epsilon_{\bm{k}}^b = \frac{\hbar^2k^2}{2m^*} - \Delta,
\end{equation}
where $m^*=0.2m_e$ is the effective electron mass, with $m_e$ the bare electron mass, and $\Delta$ is a constant energy splitting. The function $\gamma_{\bm{k}}$ represents the cubic Dresselhaus orbital coupling,
\begin{equation} \label{eq:gamma}
\gamma_{\bm{k}} = k_x(\beta-\alpha k_y^2) + ik_y(\beta-\alpha k_x^2),
\end{equation}
where $\beta$ and $\alpha$ are the linear and cubic Dresselhaus coefficients, respectively~\cite{PhysRev.100.580,Winkler2003}. Following the cubic Dresselhaus spin-orbit coupling~\cite{Krich2007}, we assume the relation
\begin{equation} \label{eq:relation_alpha_beta}
    \alpha = \beta \left( \frac{d}{\pi} \right)^2
\end{equation}
between the two coefficients. In the original context, this relationship reflects that the cubic term arises from the same linear coupling divided by the confinement-induced expectation value $\langle k_z^2\rangle=(\pi/d)^2$, with $d$ the sample thickness in the out-of-plane direction. Here, we assume that the same relationship applies to our phenomenological cubic orbital coupling.

\textit{Orbital Berry curvature.}---Applying the two-band model formalism of Ref.~\cite{Lee2025} to our model, we obtain the OBC as (a detailed derivation is presented in the Method section)
\begin{equation} \label{eq:orbital_berry_curvature}
\mathcal{O}_{z,\bm{k}}^{xy,\sigma}= - \frac{\sigma m_e\Delta^2J(\bm{k})^2}{4\hbar \left(\Delta^2+|\gamma_{\bm{k}}|^2 \right)^{5/2}},
\end{equation}
where the Jacobian factor $J(\bm{k})$ is given by
\begin{align}
J(\bm{k}) = & \; \partial_{k_x} \gamma_{\bm{k}}^{(1)} \partial_{k_y} \gamma_{\bm{k}}^{(2)} - \partial_{k_y} \gamma_{\bm{k}}^{(1)} \partial_{k_x} \gamma_{\bm{k}}^{(2)} \label{eq:jacobian_definition}  \\
= & \; \beta^2 -\beta \alpha k^2 - 3 \alpha^2k_x^2k_y^2, \label{eq:jacobian_result}  
\end{align}
with $\gamma_{\bm{k}}^{(1)} \equiv \mathrm{Re}(\gamma_{\bm{k}})$ and $\gamma_{\bm{k}}^{(2)} \equiv \mathrm{Im}(\gamma_{\bm{k}})$.

Figures~\ref{fig1}(a)--(b) show the momentum-space distribution of the OBC.
In the linear case ($\beta>0$, $\alpha=0$), the curvature exhibits a single peak at the $\Gamma$ point [Fig.~\ref{fig1}(a)].
In the cubic case ($\beta>0$, $\alpha>0$), by contrast, four additional peaks appear, related to one another by $C_4$ rotation symmetry [Fig.~\ref{fig1}(b)].
We term these four points the ``cubic Dresselhaus hot spots.''
The peak height at the hot spots is around twenty times larger than that at the $\Gamma$ point, despite the two points sharing the same Jacobian factor magnitude in the linear case, representing the dominant contribution to the OBC in the cubic case.

\begin{figure}[t!]
    \centering
     \includegraphics[width=\linewidth]{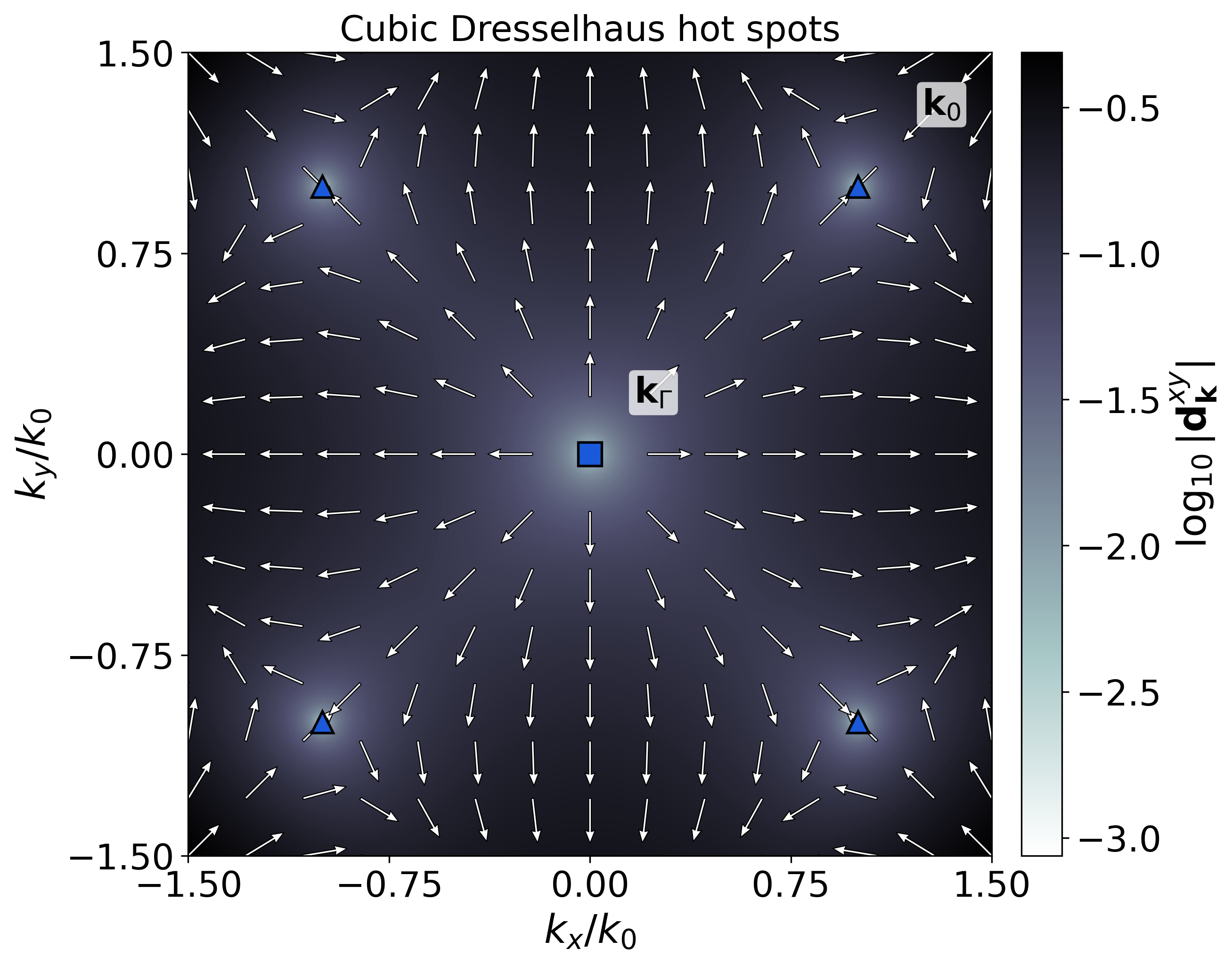}
    \caption{
    \textbf{Cubic Dresselhaus hot spots.}
    Arrows show the field direction of $\bm{d}_{\bm{k}}^{xy}=(\gamma_{\bm{k}}^{(1)},\gamma_{\bm{k}}^{(2)})$ on a regular momentum-space grid, with the background color indicating the field magnitude $|\bm{d}_{\bm{k}}^{xy}|$ (log scale). The blue square marks $\Gamma$ and the triangles mark the cubic Dresselhaus hot spots $\bm{k}_0=(\pm k_0,\pm k_0)$, where $\bm{d}_{\bm{k}}^{xy}=\bm{0}$. The parameters used are $\beta=3~\mathrm{eV\,\AA}$ and $\alpha=800\mathrm{eV\,\AA}$. 
    }
    \label{fig2}
\end{figure}

The emergence of the cubic Dresselhaus hot spots can be understood by considering the condition $\gamma_{\bm k}=0$. Neglecting the momentum dependence of the Jacobian factor, this condition minimizes the denominator of the OBC, thereby maximizing the curvature.
Solving $\gamma_{\bm k}=0$ yields the $\Gamma$ point [$\bm{k}_{\Gamma}=(0,0)$] and
\begin{equation} \label{eq:hot_spot_app}
    \bm{k}_0=(\pm k_0,\pm k_0), \qquad k_0=\sqrt{\frac{\beta}{\alpha}},
\end{equation}
which we identify as the approximate location of the hot spots.
At $\bm{k}_{\Gamma}$ and $\bm{k}_0$, the Jacobian factor is, respectively,
\begin{align}  \label{eq:jacobian_k0}
    J(\bm{k}_{\Gamma}) = \beta^2, \qquad J(\bm{k}_0) = -4\beta^2.
\end{align}
Thus, the Jacobian factor at the latter points is four times larger in magnitude, accompanied by a sign flip.
Since the OBC depends on $J(\bm{k})^2$ [see Eq.~\eqref{eq:orbital_berry_curvature}], this fourfold enhancement of $J$ translates into a sixteenfold enhancement of $\mathcal{O}_{z,\bm{k}}^{xy,\sigma}$ itself:
\begin{align} \label{eq:orbital_berry_curvature_k0}
\mathcal{O}_{z,\bm{k}=\bm{k}_\Gamma}^{xy,\pm}=\mp\dfrac{m_e\beta^4}{4\hbar\Delta^3}, \qquad \mathcal{O}_{z,\bm{k}=\bm{k}_0}^{xy,\pm}=\mp\dfrac{4m_e\beta^4}{\hbar\Delta^3}.
\end{align}
The approximate and exact hot-spot locations coincide only in the $\Delta \to 0$ limit; away from this limit, their growing separation [Fig.~\ref{fig1}(c)] causes the approximate OBC peak value to increasingly underestimate the exact one [Fig.~\ref{fig1}(d)]. Consequently, the ratio between the peak OBC values at the hot spots and at the $\Gamma$ point grows monotonically with $\Delta$ as well, increasing from $16$ in the $\Delta \to 0$ limit (consistent with Eq.~\eqref{eq:orbital_berry_curvature_k0}) to $57$ at $\Delta = 0.4$~eV.

This enhancement can be understood intuitively by considering the vector field $(d_{\bm{k}}^x, d_{\bm{k}}^y)=(\gamma_{\bm{k}}^{(1)}, \gamma_{\bm{k}}^{(2)})$.
As illustrated in Fig.~\ref{fig2}, the field forms a vortex texture around the $\Gamma$ point, whereas each hot spot forms an antivortex-type texture.
Near each singular point, the field is linearized and approximately given by
\begin{align}
    (d_{\bm{k}}^x,d_{\bm{k}}^y) & \approx \beta(k_x,k_y), & \bm{k}&\approx \bm{k}_\Gamma, \label{eq:d_vec_Gamma} \\
    (d_{\bm{k}}^x,d_{\bm{k}}^y) & \approx 2\beta\,(-q_y,-q_x), & \bm{k}&\approx \bm{k}_0, \label{eq:d_vec_hot_spot}
\end{align}
where $(q_x,q_y) \equiv (k_x-k_0, k_y-k_0)$ denotes the expansion around $\bm{k}_0$.
The cubic term cross-couples $\gamma_{\bm{k}}^{(1)}$ to $k_y^2$ and $\gamma_{\bm{k}}^{(2)}$ to $k_x^2$, effectively swapping $k_x\leftrightarrow k_y$ near the hot spots; this swap both reverses the winding number relative to the $\Gamma$ point and, upon linearization, doubles the coefficient, so that the winding rate at the hot spots is twice as fast.
Entering the Jacobian factor [Eq.~\eqref{eq:jacobian_definition}] quadratically, this twofold enhancement of the winding rate is amplified into the fourfold enhancement of $J(\bm{k})$ at the hot spots and, in turn, the sixteenfold enhancement of the OBC.

\begin{figure}[t!]
    \centering
    \includegraphics[width=\linewidth]{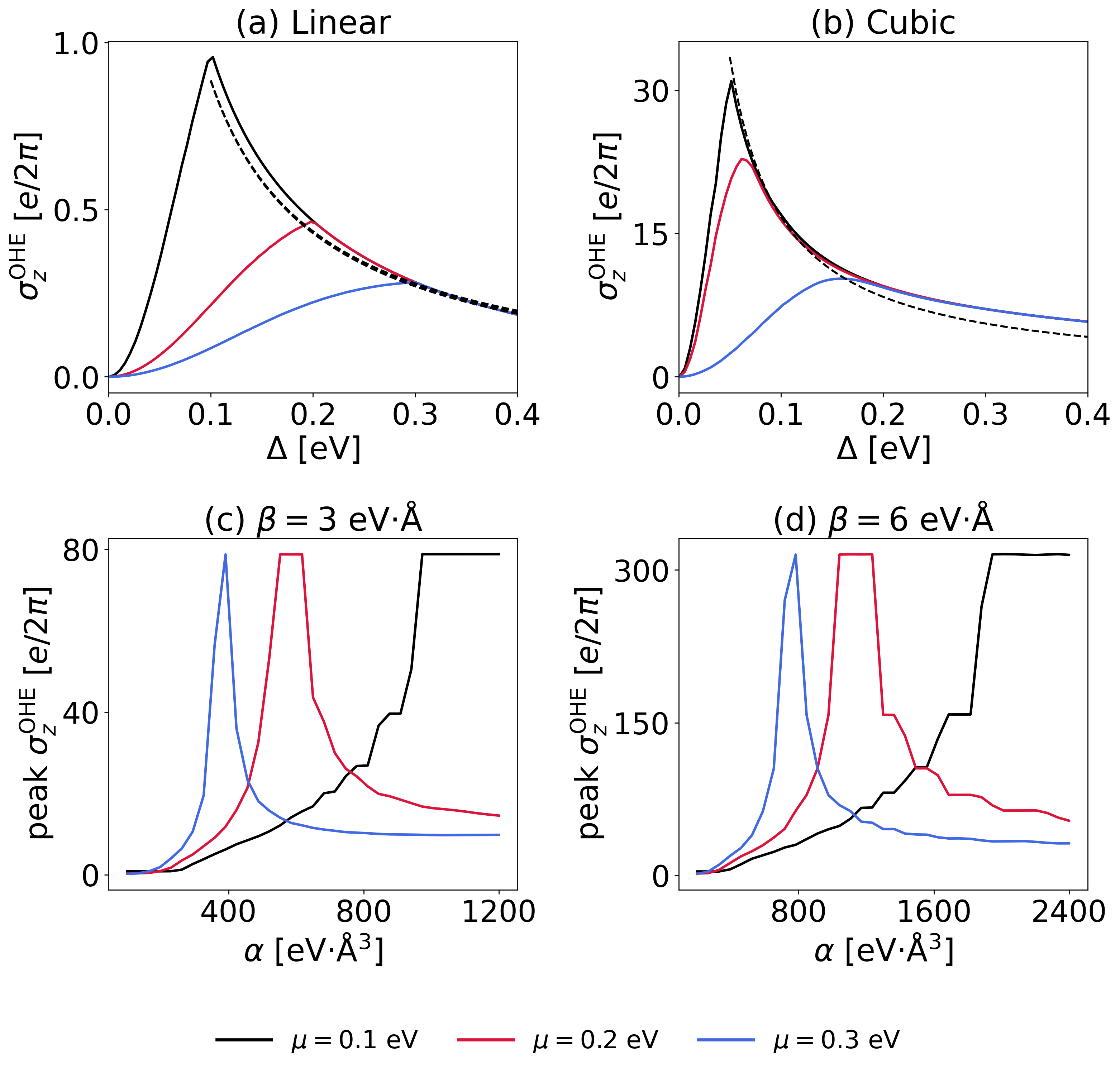}
    \caption{
    \textbf{Orbital Hall conductivity (OHC).}
    (a)--(b) OHC as a function of the energy splitting $\Delta$ for (a) the linear case ($\beta=3~\mathrm{eV\,\text{\AA}}$, $\alpha=0$) and (b) the cubic case ($\beta=3~\mathrm{eV\,\text{\AA}}$, $\alpha=800~\mathrm{eV\,\text{\AA}^3}$). Three values of the chemical potential, $\mu=0.1$, $0.2$, and $0.3$~eV, are used, as indicated by the legend. 
    Black lines show the fitting lines of Eq.~\eqref{eq:ohc_exact_lowerband} in (a), which is exact for $\Delta \geq \mu$; and Eq.~\eqref{eq:OHC_scaling_form} in (b), the small-$\Delta$ scaling form, which best approximates the numerical curve near the peak ($\Delta \approx 0.1$--$0.15~\mathrm{eV}$) and deviates progressively away from it---underestimating at very small $\Delta$ and overestimating at larger $\Delta$.
    (c)--(d) OHC as a function of $\alpha$ and chemical potential $\mu$ for (c) $\beta=3~\mathrm{eV\,\text{\AA}}$ and (d) $\beta=6~\mathrm{eV\,\text{\AA}}$, with $\Delta$ fixed along the peak condition of Eq.~\eqref{eq:popul_cond_lower_band_cubic}. 
    The temperature used is $T=0 K$ for all panels.}

    \label{fig3}
\end{figure}

\textit{Giant orbital Hall conductivity.}---The OHC is given by~\cite{Lee2025}
\begin{equation} \label{eq:orbital_hall_conductivity}
\sigma_z^{\mathrm{OHE}} = \frac{e}{\hbar}\sum_{\sigma=\pm} \int \frac{d^2\bm{k}}{(2\pi)^2} f(\epsilon_{\bm{k}}^\sigma) \mathcal{O}_{z,\bm{k}}^{xy,\sigma},
\end{equation}
where $f(E)=\dfrac{1}{\exp[(E-\mu)/k_BT]+1}$ is the Fermi--Dirac distribution function.
Using this formula, we numerically compute the OHC. Figure~\ref{fig3}(a) displays the result for the linear case with $\beta>0$ and $\alpha=0$. In the small-$\Delta$ regime ($\Delta<\mu$), the OHC initially increases with $\Delta$. This behavior can be understood by noting that both the upper ($\epsilon_{\bm{k}}^+$) and lower ($\epsilon_{\bm{k}}^-$) bands are partially occupied, and since the OBC of the two bands is exactly antisymmetric,
\begin{equation} \label{eq:26}
\mathcal{O}_{z,\bm{k}}^{xy,+} = - \mathcal{O}_{z,\bm{k}}^{xy,-},
\end{equation}
their contributions cancel over the region of momentum space where the upper band is occupied. As $\Delta$ increases, this region shrinks, reducing the cancellation and thereby increasing the OHC. At the threshold value
\begin{equation}
\Delta=\mu, \label{eq:depopul_cond_linear}
\end{equation}
the upper band becomes completely depopulated and the OHC reaches its peak. Beyond this threshold ($\Delta>\mu$), the upper band remains fully depopulated, and the behavior is governed entirely by the OBC of the lower band, which decays with $\Delta$, causing the OHC to decrease monotonically. Notably, for $T=0$, the OHC in this regime can be computed analytically (black dashed line in the figure; detailed expression presented in the Method section), and the OHC for different values of $\mu$ nearly collapses onto a single curve.

We next turn to the cubic case with $\beta,\alpha>0$ [Fig.~\ref{fig3}(b)]. 
While the OHC displays a peaking behavior with qualitatively similar to the linear case, the OHC displays a huge enhancement relative to the linear case: its peak values for $\mu=0.1$, $0.2$, and $0.3$~eV are $31.0$, $22.8$, and $10.3$ in unit of $e/(2\pi)$, respectively, each of which is $32.3$, $49.2$, and $36.3$ times larger than the corresponding peak value in the linear case [Fig.~\ref{fig3}(a)]. 
This huge enhancement can be partially explained by the growing OBC enhancement: the peak OBC ratios between the hot spots and the $\Gamma$ point at the corresponding peak $\Delta$ values are 17.0, 17.4, and 24.6 for $\mu = 0.1$, $0.2$, and $0.3$~eV, respectively [Fig.~\ref{fig1}(d)]. 
The remaining factor of 1.5 to 2.8 reflects two competing effects: the fourfold multiplicity of the cubic Dresselhaus hot spots, partially offset by their narrower individual extent in momentum space compared to the $\Gamma$ point.
These two factors associated with the cubic term compound, leading to a strong enhancement of the OHC.

Another interesting feature of the cubic case is that the OHC exhibits a more complex peaking structure than the linear case.
For the smallest $\mu$ considered ($0.1$~eV), the $\Gamma$ point of the lower band is occupied, but its cubic Dresselhaus hot spots remain unoccupied at small $\Delta$.
As $\Delta$ increases, the OHC grows until the hot spots become populated at a threshold
\begin{equation} \label{eq:popul_cond_lower_band_cubic}
\Delta  = - \mu + \frac{\hbar^2\beta}{m^*\alpha},
\end{equation}
at which point the OHC reaches a peak.
Beyond this point, a further increase of $\Delta$ decreases the OHC, since the OBC itself decreases with $\Delta$.
For the larger values of $\mu$ ($0.2$ and $0.3$~eV), by contrast, the hot spots of the lower band are already occupied at $\Delta=0$, so this mechanism is absent.
The peak position is instead determined by the depopulation of the upper band at the hot spots,
\begin{equation} \label{eq:depopul_cond_upper_band_cubic}
        \Delta = \mu - \frac{\hbar^2\beta}{m^*\alpha}.
\end{equation}
Below this threshold, the upper band's contribution---which is of opposite sign to that of the lower band---partially cancels the latter's dominant contribution.
Once $\Delta$ exceeds this threshold, the upper band depopulates and this cancellation vanishes entirely; the OHC is then governed solely by the OBC of the lower band, which decreases with $\Delta$, causing the OHC to decrease monotonically as before.
These estimates are in good agreement with the peaks observed numerically: for $\mu=0.1$, Eq.~\eqref{eq:popul_cond_lower_band_cubic} gives $\Delta=0.043$~eV, close to the numerically observed peak at $\Delta=0.052$~eV; for $\mu=0.2$ and $0.3$~eV, Eq.~\eqref{eq:depopul_cond_upper_band_cubic} gives $\Delta=0.057$~eV and $\Delta=0.157$~eV, close to the numerically observed peaks at $\Delta=0.062$~eV and $\Delta=0.163$~eV.

We now examine how the maximum achievable value of the OHC can be found. Four parameters---$\beta$, $\alpha$, $\Delta$, and $\mu$---affect the OHC, but only three can be adjusted independently, since $\Delta$ is fixed by the other three through Eq.~\eqref{eq:popul_cond_lower_band_cubic} in order to remain at the OHC peak.
Here, we therefore fix $\beta$ and vary $(\alpha,\mu)$, adjusting $\Delta$ according to Eq.~\eqref{eq:popul_cond_lower_band_cubic}.
As shown in Fig.~\ref{fig3}(c) for $\beta=3~\mathrm{eV\,\text{\AA}}$, the OHC is maximized along the curve $\mu=38.1\,(\beta/\alpha)$~eV, where the fine-tuning condition Eq.~\eqref{eq:popul_cond_lower_band_cubic} forces $\Delta\approx0$.
This can be understood from the scaling relation
\begin{equation} \label{eq:OHC_scaling_form}
    \sigma_z^{\rm OHE} \simeq 17\,\dfrac{em_e\beta^2}{24\pi\hbar^2}\,\Delta^{-1},
\end{equation}
where, among the factor of $17$, $16$ arises from the contribution of the hot spots and $1$ from that of the $\Gamma$ point; see the Methods section for the derivation of this scaling.
The largest OHC is therefore obtained by simultaneously satisfying the population condition Eq.~\eqref{eq:popul_cond_lower_band_cubic} while minimizing $\Delta$, which is achieved in Fig.~\ref{fig3}(c) by tuning $\alpha$.
It is worth noting that the divergent behavior with $\Delta$ contrasts with that of the ordinary Berry curvature: once the occupied Fermi sea contains the singular point, its integrated contribution instead approaches a $\Delta$-independent value~\cite{PhysRevLett.99.236809}.
This qualitative difference originates from the extra factor $\Delta(\Delta^2+|\gamma_{\bm{k}}|^2)^{-1}$ that the OBC carries relative to the ordinary Berry curvature [cf.\ Eqs.~\eqref{eq:orbital_berry_curvature} and~\eqref{eq:berry_curvature}]. 
This factor sharpens the singularity at $\bm{k}_s$ and prevents the cancellation between the enhanced peak height and shrinking peak area that otherwise yields a constant value under the momentum integral.
For a larger value of $\beta=6~\mathrm{eV\,\text{\AA}}$, the behavior of $\sigma_z^{\rm OHE}$ remains qualitatively similar [Fig.~\ref{fig3}(d)], suggesting that the dependence on $\beta$ is comparatively simple: it scales approximately as $\sigma_z^{\mathrm{OHE}}\propto\beta^2$, with deviations arising because $\beta$ also enters the denominator of the OBC through $\gamma_{\bm{k}}$.

\begin{figure} [t!]
    \centering
    \includegraphics[width=\linewidth]{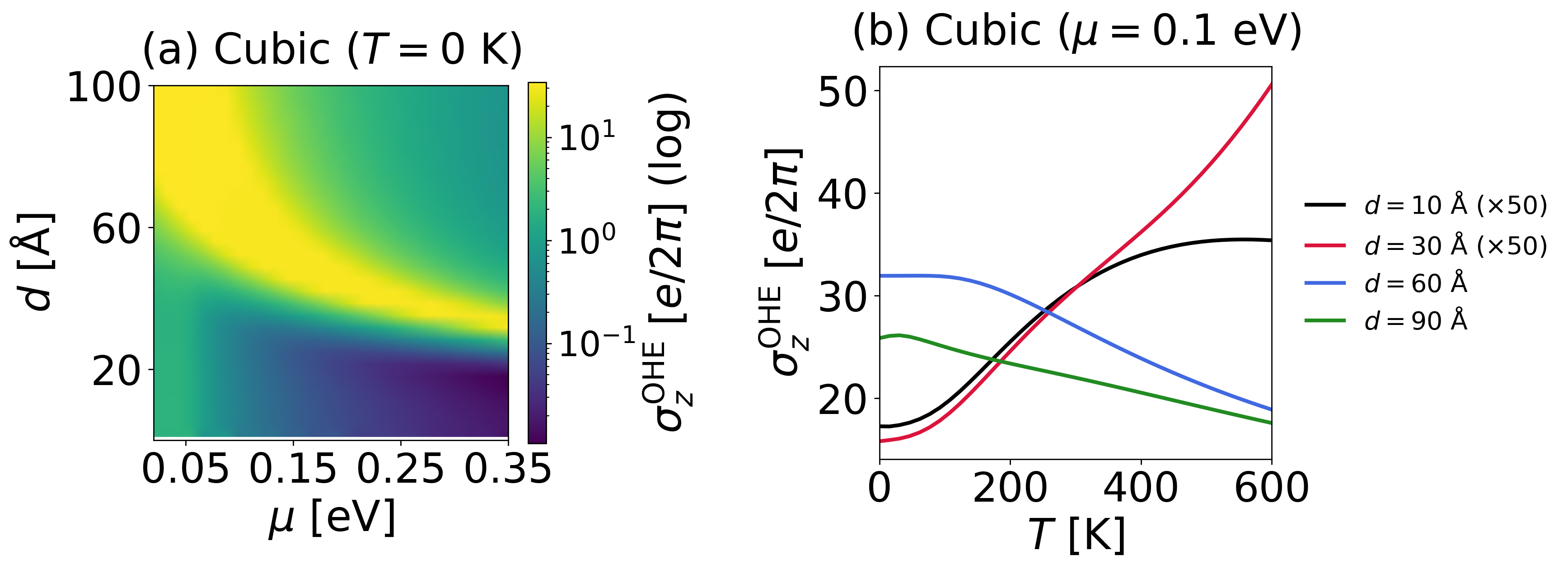}
    \caption{
    \textbf{Experimental signatures.}
    (a) OHC as a function of chemical potential $\mu$ and sample thickness $d$, at $T=0$.
    (b) OHC as a function of temperature $T$ for different values of $d$, at $\mu=0.1$~eV.
    Both panels use $\beta=3~\mathrm{eV\,\AA}$ and $\Delta=0.05$~eV
    }
    \label{fig4}
\end{figure}

\textit{Experimental signatures.}---Combining the relationship $\alpha=\beta(d/\pi)^2$ with the hot-spot location $k_0=\sqrt{\beta/\alpha}$ yields
\begin{equation} \label{eq:relation_k_0_d}
    k_0 = \frac{\pi}{d},
\end{equation}
suggesting that the hot spot location -- and hence the OHC -- is tunable via the sample thickness $d$. Figure~\ref{fig4}(a) presents the OHC as a function of $d$ and $\mu$: the OHC is strongly enhanced only when $d$ and $\mu$ are jointly tuned so that the hot spots fall within the occupied Fermi sea, giving rise to a sharply peaked ridge in the $(d,\mu)$ plane rather than a broad enhancement. Figure~\ref{fig4}(b) presents the OHC as a function of $d$ and $T$, revealing two qualitatively distinct thermal behaviors depending on whether the hot spots are occupied at $T=0$.
For intermediate $d$, where the lower-band hot spots are already occupied at $T=0$, raising the temperature thermally populates the upper band, whose OBC is of opposite sign, so the OHC is progressively suppressed with increasing $T$.
For small $d$, by contrast, the hot spots lie above $\mu$ and remain unoccupied at $T=0$; here, thermal excitation instead populates the lower-band hot spots themselves, so the OHC increases with $T$.
This non-monotonic, $d$-dependent thermal response provides a decisive experimental signature of the cubic Dresselhaus orbital coupling proposed in this work, distinguishing it from the purely linear case.

\textit{Discussion.}---In summary, we showed that the cubic Dresselhaus orbital coupling gives rise to a giant OHE. This enhancement stems from two effects: the enhanced OBC at the cubic Dresselhaus hot spots, and the multiplicity of the hot spots themselves. Realizing this coupling requires a host system with (i) a pair of orbitally degenerate or near-degenerate bands to serve as the pseudospin $(a,b)$, and (ii) bulk inversion asymmetry of the zinc-blende symmetry class, which permits the cubic-in-momentum coupling of Eq.~\eqref{eq:gamma}. Identifying specific candidate materials is left for future work. The thickness, chemical-potential, and temperature dependences of the OHC established above can serve as experimental signatures of this effect.

The underlying mechanism -- steepened phase winding near momentum-space hot spots -- is not specific to the OHE. We already showed the Berry curvature exhibits a fourfold enhancement at the hot spots (and quantum metric as well; see Supplementary Note~3), suggesting a broader role for hot-spot physics in related quantum transport phenomena. Given the functional form of these quantities' dependence on the level splitting, we predict that transport phenomena based on them acquire a universal fourfold enhancement from the cubic coupling in the small-splitting regime. More broadly, identifying such higher-order hot spots in other types of higher-order spin or orbital coupling may prove useful~\cite{Studer2010}. We particularly note that light-induced cubic Rashba coupling~\cite{yvvk-9b74} and even-order intersubband coupling~\cite{hk72-nyhz} offer promising settings for analogous enhancements.

\section*{Methods}

\textbf{Derivation of the OBC.}---The Berry curvature is given by~\cite{Lee2025}
\begin{equation} \label{eq:berry_curvature_theta_zeta}
\Omega_{\bm{k}}^{ij,\sigma} = \frac{\sigma}{2\zeta_{\bm{k}}^2}
\left( \frac{\partial \zeta_{\bm{k}}}{\partial k_i} \frac{\partial \theta_{\bm{k}}}{\partial k_j} - \frac{\partial \zeta_{\bm{k}}}{\partial k_j} \frac{\partial \theta_{\bm{k}}}{\partial k_i} \right),
\end{equation}
where $\theta_{\bm{k}}$ is the phase of the interorbital coupling and $\zeta_{\bm{k}}$ is an amplitude factor entering the eigenstates
\begin{equation} \label{eq:6}
    |\psi_{\bm{k}}^{\sigma}\rangle = \frac{1}{\sqrt{2\zeta_{\bm{k}}}}\left(\sigma e^{i\theta_{\bm{k}}}\sqrt{\zeta_{\bm{k}}+\sigma}, \quad \sqrt{\zeta_{\bm{k}}+\bar{\sigma}}\right)^{\mathrm{T}},
\end{equation}
with $\bar{\sigma}=\mp$ for $\sigma=\pm$; $\theta_{\bm{k}}$ and $\zeta_{\bm{k}}$ are defined respectively as
\begin{equation} \label{eq:theta_zeta}
    \theta_{\bm{k}}=\arg(\gamma_{\bm{k}}), \qquad \zeta_{\bm{k}}=\sqrt{1+\frac{|\gamma_{\bm{k}}|^2}{\Delta^2}}.
\end{equation}
Substituting Eq.~\eqref{eq:theta_zeta} into Eq.~\eqref{eq:berry_curvature_theta_zeta}, the Berry curvature can be written directly in terms of $\gamma_{\bm{k}}$ and $\Delta$ as
\begin{equation} \label{eq:berry_curvature}
    \Omega_{\bm{k}}^{ij,\sigma} = \frac{\sigma\Delta J(\bm{k})}{2\left(\Delta^2+|\gamma_{\bm{k}}|^2\right)^{3/2}}.
\end{equation}
where the Jacobian factor $J(\bm{k})$ is provided in Eq.~\eqref{eq:jacobian_result}.
Using the eigenvalues of $H_{\bm{k}}$
\begin{equation} \label{eq:eigenvalues}
    \epsilon_{\bm{k}}^{\sigma} = \frac{\hbar^2k^2}{2m^*} +\sigma\sqrt{\Delta^{2}+|\gamma_{\bm{k}}|^{2}},
\end{equation}
the band splitting evaluates to
\begin{equation} \label{eq:band_splitting}
\epsilon_{\bm{k}}^+ - \epsilon_{\bm{k}}^- = 2\sqrt{\Delta^2+|\gamma_{\bm{k}}|^2}.
\end{equation}
Substituting this expression and Eq.~\eqref{eq:berry_curvature} into the OBC expression~\cite{Lee2025},
\begin{equation} \label{eq:orbital_berry_curvature_formula}
\mathcal{O}_{z,\bm{k}}^{ij,\sigma} = -\frac{\sigma m_e}{4\hbar} \left(\epsilon_{\bm{k}}^+ - \epsilon_{\bm{k}}^- \right) \left( \Omega_{\bm{k}}^{ij,\sigma} \right)^2,
\end{equation}
yields the result given in Eq.~\eqref{eq:orbital_berry_curvature}; a more detailed derivation is provided in Supplementary Note~1.

\textbf{Precise hot-spot location.}---Differentiating Eq.~\eqref{eq:orbital_berry_curvature} with respect to $k_x$ and $k_y$ and setting the result to zero gives
\begin{equation} \label{eq:hot_spot_precise}
\begin{aligned}
F(k_x,k_y) ={}&5\beta^4+ \alpha\beta\left(4\Delta^2- \beta^2 k_x^2- 21\beta^2 k_y^2\right)\\
&+ \alpha^2\left[12\Delta^2 k_y^2+ 11\beta^2 k_x^2 k_y^2+ 37\beta^2 k_y^4\right]\\
&+ \alpha^3\beta\left[k_x^2 k_y^4- 6k_x^4 k_y^2- 5k_y^6\right]\\
&- \alpha^4\left[3k_x^2 k_y^6+ 18k_x^4 k_y^4\right]=0.
\end{aligned}
\end{equation}
with the corresponding equation for $\partial_{k_y}\mathcal{O}_{z,\bm{k}}^{xy,\sigma}=0$ obtained by the interchange $k_x\leftrightarrow k_y$.
A detailed derivation is provided in Supplementary Note~2.
Since $\mathcal{O}_{z,\bm{k}}^{xy,\sigma}$ is invariant under $k_x\leftrightarrow k_y$, both equations are satisfied simultaneously along the diagonal $k_x=k_y$, so the two-dimensional problem reduces to the single condition $F(k,k)=0$.
We numerically confirmed that the smallest positive solution of this equation agrees with $\bm{k}_0$ [Eq.~\eqref{eq:hot_spot_app}] over the full range of parameters used: $\beta=3,6~\mathrm{eV\,\text{\AA}}$; $\alpha=0,800~\mathrm{eV\,\text{\AA}^3}$; and $\Delta<0.4~\mathrm{eV}$.


\textbf{Exact OHC result for the linear case.}---For the linear case ($\alpha=0$) at $T=0$ and $\Delta>\mu$, the OHC integral reduces to
\begin{equation} \label{eq:ohc_integral_linear}
    \sigma_z^{\mathrm{OHE}} = \frac{e}{\hbar}\int_0^{k_F} \frac{k\,dk}{2\pi} \frac{m_e\Delta^2\beta^4}{4\hbar(\Delta^2+\beta^2k^2)^{5/2}},
\end{equation}
where $k_F$ is the Fermi wave vector of the lower band,
\begin{equation} \label{eq:kF_exact}
    k_F = \sqrt{\frac{2A\mu+\beta^2+\sqrt{4A^2\Delta^2+4A\beta^2\mu+\beta^4}}{2A^2}},
\end{equation}
with $A=\dfrac{\hbar^2}{2m^*}$.
This integral can be evaluated analytically in closed form,
\begin{equation} \label{eq:ohc_exact_lowerband}
    \sigma_z^{\mathrm{OHE}} = \frac{e\,m_e\,\beta^2}{24\pi\hbar^2}\left[\Delta^{-1} - \Delta^2\left(\Delta^2+\beta^2 k_F^2\right)^{-3/2}\right].
\end{equation}
The black curve in Fig.~3(a) is obtained by plotting this formula, multiplied by an empirical factor of $0.90$ (accounting for the finite $k_{\max}$ integration box), over the range $0.1~\mathrm{eV} \leq \Delta \leq 0.4~\mathrm{eV}$.

\textbf{Scaling argument for the $1/\Delta$ divergence of OHC.}---To derive Eq.~\eqref{eq:OHC_scaling_form}, we linearize the coupling near a singular point $\bm{k}_s$ ($\bm{k}_\Gamma$ or $\bm{k}_0$) where $\gamma_{\bm{k}_s}=0$, taking $|\gamma_{\bm{k}}| \approx \sqrt{|J(\bm{k}_s)|}\,|\delta\bm{k}|$ with $\delta\bm{k}=\bm{k}-\bm{k}_s$ [Eqs.~\eqref{eq:d_vec_Gamma}--\eqref{eq:d_vec_hot_spot}], where $J(\bm{k})\approx J(\bm{k}_s)\equiv J_s$ is approximately constant over the relevant range of $\delta\bm{k}$.
Substituting this linearized form into the OBC [Eq.~\eqref{eq:orbital_berry_curvature}] and extending the resulting integral over $\delta\bm{k}$ to the full plane (justified since the integrand decays rapidly away from $\bm{k}_s$), the total contribution of all singular points to the OHC integral [Eq.~\eqref{eq:orbital_hall_conductivity}] can be evaluated exactly:
\begin{equation}
    \sigma_z^{\rm OHE} \simeq \sum_{\bm{k}_s}\frac{e}{\hbar}\int \frac{d^2(\delta\bm{k})}{(2\pi)^2}\, \frac{m_e\Delta^2J_s^2}{4\hbar\left(\Delta^2+J_s|\delta\bm{k}|^2\right)^{5/2}},
\end{equation}
where the sum over $\bm{k}_s$ runs over the singular points, $\bm{k}_\Gamma$ or $\bm{k}_0$.
Carrying out the integral gives
\begin{equation}
    \sigma_z^{\rm OHE}= \sum_{\bm{k}_s}\frac{em_e\Delta^2J_s^2}{4\hbar^2}\cdot\frac{1}{6\pi J_s\Delta^3} = \sum_{\bm{k}_s}\frac{em_e}{24\pi\hbar^2}\,\frac{J_s}{\Delta},
\end{equation}
where the last equality follows from $\int_0^\infty \pi\,du\,(\Delta^2+J_su)^{-5/2}=2\pi/(3J_s\Delta^3)$ with $u=|\delta\bm{k}|^2$.
It is crucial to note that, although the OBC itself scales as $J_s^2$, the integrated OHC depends on $J_s$ only linearly: the $J_s^2$ peak height is partly offset by the shrinking peak area, $\delta k^2\sim\Delta^2/J_s$, which scales as $J_s^{-1}$, leaving a net $J_s^2\times J_s^{-1}=J_s$ dependence.
Both the $\Gamma$ point and the four hot spots contribute, with $J_\Gamma=\beta^2$ and $J_{\bm{k}_0}=4\beta^2$ at each of the four hot spots, so that
\begin{equation}
    \sigma_z^{\rm OHE}\Big|_{\rm cubic} \simeq \frac{em_e}{24\pi\hbar^2\Delta}\left[\beta^2 + 4\times(4\beta^2)\right],
\end{equation}
where the first and second terms originate from the $\Gamma$ point and the four hot spots, respectively.
This yields the scaling form of the OHC quoted in Eq.~\eqref{eq:OHC_scaling_form}, and the black curve in Fig.~\ref{fig3}(b) is obtained by plotting this formula. For the linear case, only the $\Gamma$ point contributes, with $J_\Gamma=\beta^2$, giving
\begin{equation}
    \sigma_z^{\rm OHE}\Big|_{\rm linear} \simeq \frac{em_e\beta^2}{24\pi\hbar^2}\,\Delta^{-1},
\end{equation}
which reproduces exactly the $\Delta^{-1}$ term of Eq.~\eqref{eq:ohc_exact_lowerband}.

\begin{acknowledgements}
This research was supported by an appointment to the JRG Program at the APCTP through the Science and Technology Promotion Fund and Lottery Fund of the Korean Government, the Korean Local Governments (Gyeongsangbuk-do Province and Pohang City), and the National Research Foundation of Korea (NRF) funded by the Korean government (Ministry of Science and ICT, MSIT) (No. RS-2026-25499525).
\end{acknowledgements}

\bibliography{ref}

\end{document}